# Collective magnetism at multiferroic vortex domain walls

Yanan Geng, N. Lee, Y.J. Choi[*], S-W. Cheong and Weida Wu

*Department of Physics and Astronomy and Rutgers Center for emergent materials, Rutgers University, Piscataway, NJ 08854 USA*

Topological defects have been playgrounds for many emergent phenomena in complex matter such as superfluids, liquid crystals, and early universe. Recently, vortex-like topological defects with six interlocked structural antiphase and ferroelectric domains merging into a vortex core were revealed in multiferroic hexagonal manganites. Numerous vortices are found to form an intriguing self-organized network. Thus, it is imperative to find out the magnetic nature of these vortices. Using cryogenic magnetic force microscopy, we discovered unprecedented alternating net moments at domain walls around vortices that can correlate over the entire vortex network in hexagonal $ErMnO_3$. The collective nature of domain wall magnetism originates from the uncompensated $Er^{3+}$ moments and the correlated organization of the vortex network. Furthermore, our proposed model indicates a fascinating phenomenon of field-controllable spin chirality. Our results demonstrate a new route to achieving magnetoelectric coupling at domain walls in single-phase multiferroics, which may be harnessed for nanoscale multifunctional devices.

Multiferroics are materials with coexisting magnetism and ferroelectricity[1]. The cross-coupling between two ferroic orders can result in giant magnetoelectric coupling for potential applications[2-5]. Because formation of domains is the hallmark of any ferroic order[6], it is of both fundamental and technological interests to visualize cross-coupled domains or walls in multiferroics. Hexagonal (*h*-) *RE*MnO$_3$ (*RE* = Sc, Y, Ho, … Lu) are multiferroics with coexistence of ferroelectricity ($T_C \approx 1200 - 1500$ K)[7] and antiferromagnetism ($T_N \approx 70 - 120$ K)[8]. The ferroelectricity is induced by structural instability called trimerization[9,10], which lifts presumably the frustration of antiferromagnetic interactions of Mn$^{3+}$ spins on triangular lattice. Indeed, a 120º antiferromagnetic order of Mn$^{3+}$ spin in the *ab*-plane sets in below $T_N$. Recently, an intriguing 6-state vortex domain structure in YMnO$_3$ is revealed by transmission electron microscopy, conductive atomic force microscopy and piezoresponse force microscopy (PFM) at room temperature[11-13]. The formation of 6-state vortices originates from the cyclic arrangement of 6 interlocked structural antiphase (α, β, γ) and ferroelectric (+/−) ground states (i.e. α$^+$, β$^-$, γ$^+$, α$^-$, β$^+$, γ$^-$)[11,14]. The intriguing network of vortex-antivortex pairs has a profound connection to graph theory, where 6-valent planer graphs with even-gons are two-proper-colorable[15]. Using second harmonic generation optics, it has been claimed that ferroelectric domain walls (DWs) in millimeter-size YMnO$_3$ tend to pin antiferromagnetic DWs, but free antiferromagnetic DWs also exist[16]. Thus, it is of great interest to find out the magnetic nature of vortex domains and DWs. However, this has been an experimental challenge, particularly due to the lack of suitable high resolution imaging technique of antiferromagnetic domains or DWs for *h-RE*MnO$_3$ at low temperatures (supplementary discussion 1). To address this vital issue, we visualized magnetic structure of *h*-ErMnO$_3$ crystals using a low-temperature magnetic force microscope[17].

Herein, we report discovery of remarkable alternating net moments at interlocked antiphase-ferroelectric DWs around vortex cores in multiferroic *h*-ErMnO$_3$. Our results suggest that the intriguing DW magnetism originates from uncompensated Er$^{3+}$ spins polarized by the Mn$^{3+}$ antiferromagnetic order via Dzyaloshinskii-Moriya (DM) interaction[18,19]. More interestingly, the DW net moments correlates over the entire vortex network and can be controlled by magnetic field, suggesting a collective nature of the DW magnetism. The net magnetic moments at ferroelectric DWs may potentially be manipulated by applied electric field, indicating an intriguing nanoscale magnetoelectric coupling at nanoscale. In addition, our simple model of DW magnetism indicates an extraordinary field-controllable single chiral spin state.

**Results**

Figure 1a shows a room temperature PFM image taken on the (001) surface of a *h*-ErMnO$_3$ single crystal, where six alternating up (red) and down (blue) ferroelectric domains merge into a vortex core. Fig. 1b and 1c show MFM images measured at 5.5 K in 0.2 T out-of-plane (OOP) magnetic field after -0.2 T (+0.2 T) field cooling from 100 K (>$T_N$). The MFM images were taken on the *same* location as PFM image (fig. 1a). Fig. 1d shows the MFM measurement setup, where 50 nm Au film was deposit on sample surface to eliminate electrostatic stray field from OOP ferroelectric domains. The MFM tip moment is normal to sample surface so that the MFM signal (cantilever resonant frequency shift $\Delta f \propto$ force gradient) is due to OOP stray magnetic field gradient from sample[20].

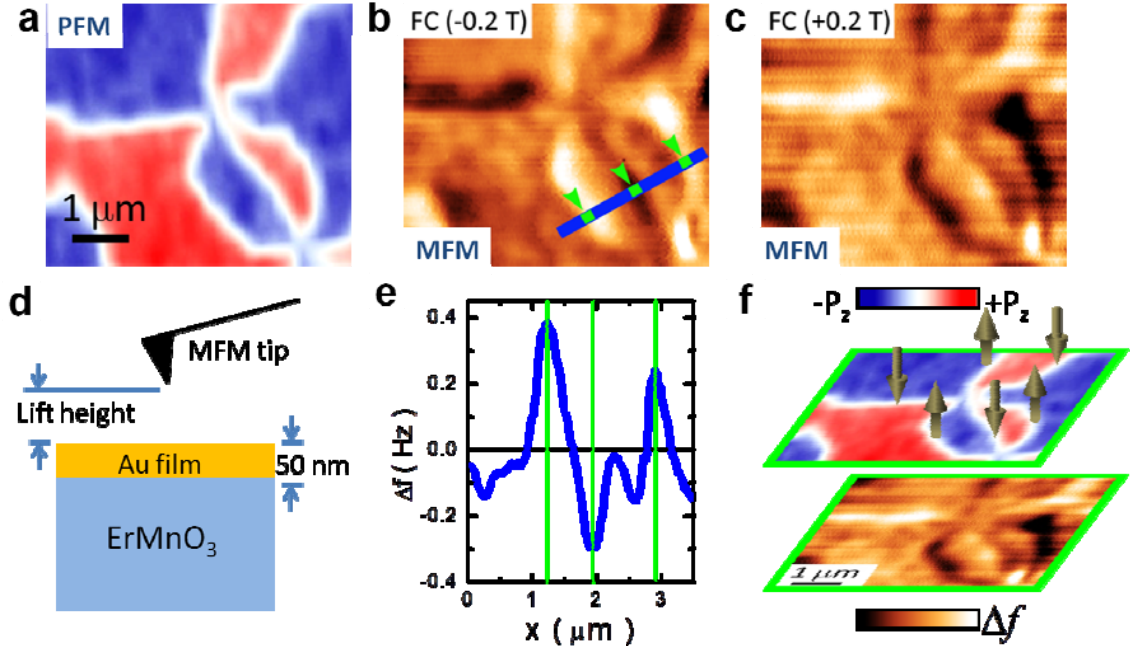

**Figure 1 | Coupled antiphase-ferroelectric and antiferromagnetic domain walls with alternating magnetic moments around mutliferroic vortex cores. a,** room temperature PFM image on the (001) surface of a *h*-ErMnO$_3$ single crystal. The red and blue colors in PFM image correspond to up and down ferroelectric domains, respectively. **b** (**c**), MFM image measured at 5.5 K in 0.2 T OOP magnetic field after -0.2 T (+0.2 T) field cooling from 100 K (>$T_N$) to 5.5 K. MFM images were taken at the *same* location as the PFM image, The color scale ($\Delta f$) is 0.8 Hz and the lift height is 50 nm. **d**, a cartoon sketch shows the setup of the MFM experiment (see method for details). **e**, the line profile of MFM signal along the blue line in **b**, vertical green lines note the position of DWs as indicated by the green arrows in **b**. **f**, Perspective view of PFM (**a**) and MFM (**c**) images with arrows representing the orientation of uncompensated magnetic moments at structural antiphase-ferroelectric DWs.

It is clear that there are remarkable line features with alternating bright and dark colors in MFM images (fig. 1b and 1c), correlating with antiphase-ferroelectric DWs around the vortex core in the PFM image (fig. 1a). The MFM signal along a line drawn in fig. 1b is shown in fig. 1e. The single peak profile (width ~400 nm) of MFM signal suggests that local magnetization at DWs is parallel to the *c*-axis. The MFM contrast of DWs is essentially constant in 0.02 – 0.7 T (see supplementary fig. S1), which excludes the possibility of local susceptibility differences as the origin[21]. Therefore, the MFM signal at DWs originates from local net moments along the *c*-axis. Since Mn$^{3+}$ spins in ErMnO$_3$ are 120º-ordered below $T_N \approx$ 80 K, no net moment is expected inside antiferromagnetic domains[22]. Naturally, these net moments at antiphase-ferroelectric DWs come from the uncompensated moments at antiferromagnetic domain boundaries. Interestingly, there is negligible MFM contrast at the vortex core, indicating that alternating net moments from 6 DWs cancel out each other at the core. Because 0.2 T OOP magnetic field is much larger than the coercive field of our MFM tip (~0.02 T), the contrast inversion between fig. 1b and 1c (both were measured at 0.2 T) is due to reversal of local net moments at DWs. The fact that cooling magnetic field determines the magnetic state of alternating vortex DWs suggests the DW magnetism in $h$-ErMnO$_3$ is likely a correlated phenomenon, presumably tied to the antiferromagnetic order of multiferroic domains. Therefore, our results provide compelling evidence that correlated net moments are coupled to the interlocked structural antiphase and ferroelectric DWs interconnecting vortices in ErMnO$_3$, which are truly multiferroic. As summarized by the perspective view of PFM and MFM images in fig. 1f, the uncompensated moments at DWs around a vortex core are parallel to the $c$-axis with alternating orientation, similar to alternating ferroelectric polarization of domains around a vortex core, suggesting that there are two types of DWs, denoted as DW$_I$ and DW$_{II}$.

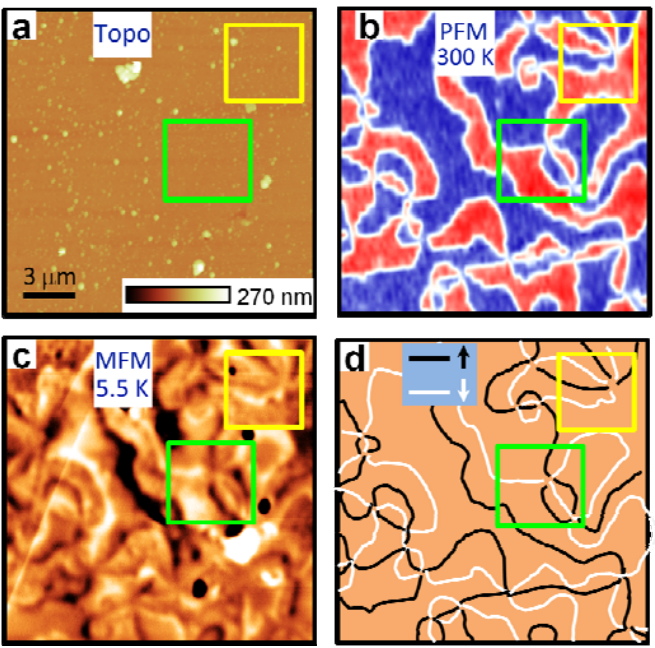

**Figure 2 | Correlation of DW magnetism over the vortex network. a,** topography **b**, room temperature PFM image, and **c**, MFM image (5.5 K, 0.2 T, lift height: 180 nm) taken at the *same* location on the (001) surface of a $h$-ErMnO$_3$ single crystal. The color scale for the topography and MFM ($\Delta f$) are 270 nm and 0.4 Hz, respectively. The red and blue colors in the PFM image correspond to up and down ferroelectric domains, respectively. **d**, a cartoon sketch of the DW net moments over the entire field of view based on MFM data in fig. 1c (the same location as the green boxs) and the vortex connectivity in **b** (PFM). Black (white) lines represent up (down) net moments.

To determine whether the correlated DW magnetism is a local of individual vortices or a global property of the vortex network, we had obtained room temperature PFM and low temperature MFM images on the *same* area (18×18 μm$^2$) using topographic features (in fig. 2a) as alignment mark, as shown in fig. 2b and 2c, respectively. To avoid tip crashes during MFM scan over tall topographic features (>100 nm), we used a larger lift height (180 nm), which reduces the spatial resolution[20]. Fig. 2d shows a cartoon of the DW net moments based on the vortex connectivity in PFM (fig. 2b) and the MFM signals of 6 DWs around the vortex in the green box (the same as fig. 1c). Except some random features that correlate with particles or topographic roughness, the DW signals in the MFM image is in excellent agreement with that in the cartoon (an example is highlighted in yellow boxes) suggesting that the alternating DW net moments correlate over the entire field of view (and possibly over the whole sample), i.e. a collective phenomenon.

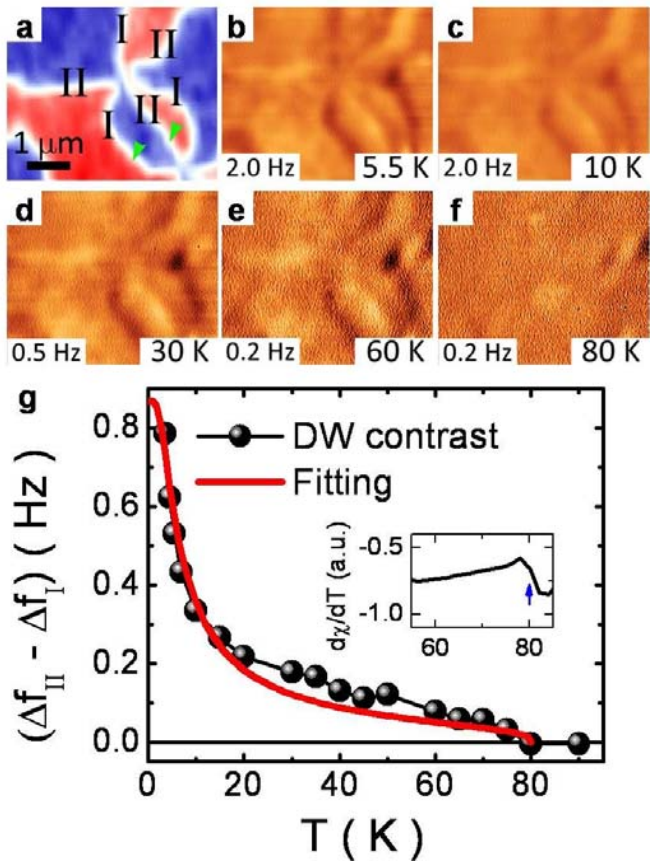

**Figure 3 | Temperature dependence of the net moments at antiphase-ferroelectric DWs. a**, PFM image at the same location as the MFM images, DW$_I$ (DW$_{II}$) are labeled by 'I' ('II'). **b-f**, selected MFM images at various temperatures (warming) in 0.2 T magnetic field. (See supplementary fig. S2 for the complete data set) The color scales are noted at the left bottom of each image. **g**, Temperature dependence of the DW contrast ($\Delta f_{II}$ - $\Delta f_I$) of two DWs noted by green arrows in **a**. The red curve is a fitting curve based on a doublet model. The inset shows the

derivative of DC susceptibility respect to temperature, d$\chi_{DC}$/d$T$. The blue arrow indicates the anomaly at $T_N$.

Previous micromagnetic analysis of antiferromagnetic DWs in *h*-YMnO$_3$ suggests oscillatory uncompensated magnetization of Mn$^{3+}$ spins across the antiferromagnetic-ferroelectric DW in the *ab*-plane due to in-plane anisotropy of Mn$^{3+}$ spins[23], which is inconsistent with our MFM results of *h*-ErMnO$_3$. Therefore, the observed uncompensated moments in ErMnO$_3$ likely come from Er$^{3+}$ spins. To confirm it, we took MFM images at the same location for various temperatures, as shown in fig. 3 (see supplementary fig. S2 for the complete data set). The DW contrast, defined as the difference between bright and dark DWs, decreases sharply for $T < 20$ K, then slowly at higher temperature as shown in fig. 3g, resembling the Curie-Weiss behavior. Assuming the MFM signal is proportional to the size of net moments, we obtained a good fit (red curve in fig. 3g) of the temperature dependence of DW contrast using a phenomenological model where the effective doublet ground state of Er$^{3+}$ is split by antisymmetric exchange fields from neighboring Mn$^{3+}$ spins[24] (supplementary discussion 2). It is believed that the effective exchange fields originate from DM interactions because dipolar interactions from Mn$^{3+}$ moments are too weak[24]. However, dipolar interactions between *RE*$^{3+}$ ions may be responsible for the additional *RE*$^{3+}$ ordering below 10 K[22,24]. The doublet model has been successful on explaining bulk (i.e. domains) *RE*$^{3+}$ ordering in other *h-RE*MnO$_3$ (*RE*=Ho, Tm, Yb) probed by x-ray magnetic resonant scattering, neutron diffraction and Mössbauer spectroscopy[24-26]. The good fitting of our MFM data demonstrates that the doublet model with DM interactions is also responsible for the remarkable DW magnetism which is invisible to most bulk probes. Indeed, the DW contrast of MFM data diminishes at 80 K, in excellent agreement with $T_N \approx 80$ K inferred from the bulk susceptibility data (inset of fig. 3g). Therefore, the DW net moments are intimately connected to the bulk antiferromagnetism of Mn$^{3+}$ spins.

**Discussion**

The DM interaction is the key ingredient for non-collinear (spiral) spin orders which break inversion symmetry and induces ferroelectricity in many recently discovered multiferroics[1,27]. In mutliferroic *h-RE*MnO$_3$, DM interactions between *RE*$^{3+}$ and Mn$^{3+}$ spins are responsible for inducing the partial *RE*$^{3+}$ antiferromagnetic order[24-26]. Therefore, the magnetic symmetry of *RE*$^{3+}$ order follows that of Mn$^{3+}$ order. In zero or low magnetic field, the magnetic space group of 120º order of Mn$^{3+}$ spins in ErMnO$_3$ is *P6$_3$cm*, i.e. B$_2$ (or Γ$_4$) in one dimensional irreducible representation[22,28]. Indeed, the Er$^{3+}$ order moments polarized by DM exchange fields also respect B$_2$ symmetry (supplementary discussion 2 and fig. S3). Earlier studies of vortices in *h*-YMnO$_3$ suggest that there may be two types of interlocked structural antiphase-ferroelectric DWs which alternate around vortex cores[11]. Assuming atomically sharp Mn$^{3+}$ spin variation at DWs, single Mn$^{3+}$ spin chirality in domains across DWs, and abrupt structural distortion variation across the two types of DWs, we are able to obtain opposite uncompensated Er$^{3+}$ moments polarized by DM exchange fields at the two types of DWs (supplementary fig. S4 and discussion 3). The result of uncompensated $Er^{3+}$ moments at $DW_I$ ($DW_{II}$) is shown in fig. 4b (4c), and is simplified in fig. 4d. In other words, two types of magnetic DWs originate from two types of interlocked structural antiphase-ferroelectric DWs assuming single $Mn^{3+}$ spin chirality. Therefore, our MFM results provide strong evidence of the existence of only 2 types of antiphase-ferroelectric DWs in h-$RE$MnO$_3$. Since the uncompensated moments couple to antiphase-ferroelectric DWs, they would likely follow the motion of DWs in the presence of applied electric field, i.e. a local magnetoelectric coupling.

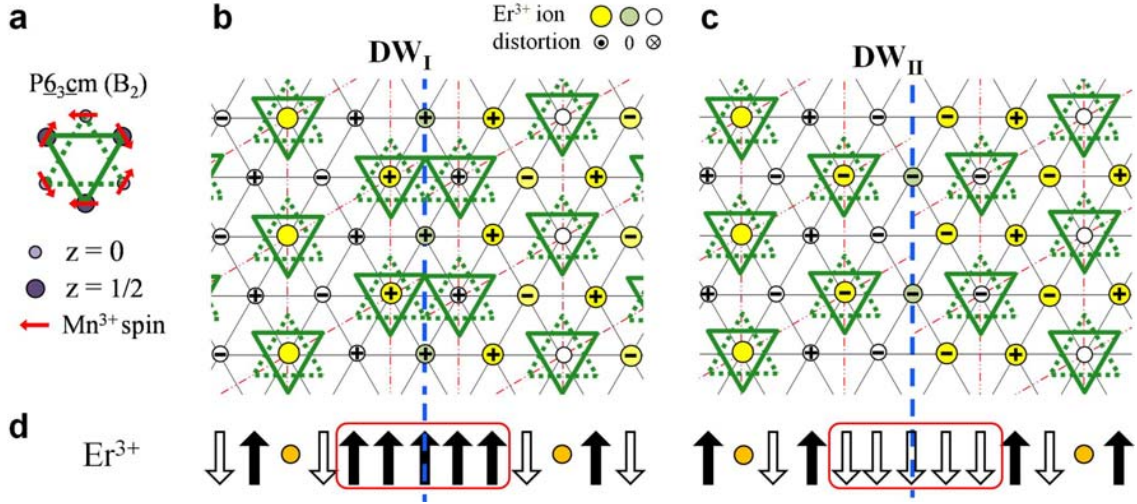

**Figure 4 | Uncompensated $Er^{3+}$ moments at DW's in $B_2$ phase. a**, $Mn^{3+}$ spins configuration of $B_2$ magnetic symmetry. Solid (dotted) green triangles represent $Mn^{3+}$ trimers at z = c/2 (z = 0) layer. $Mn^{3+}$ ions and spins are not shown in the rest cartoons for clarity. **b** and **c**, local distortion and spin configuration of $Er^{3+}$ ions in type I and type II DWs in ErMnO$_3$, respectively. Yellow (white) circle corresponds to $Er^{3+}$ ion distorting out of (into) paper. '+' ('−') denotes the induced $Er^{3+}$ moment is out of (into) paper. **d**, Cartoon of spin configuration of $Er^{3+}$ atomic planes in **b** and **c** near $DW_I$ and $DW_{II}$, the arrows inside the red boxes represent uncompensated moments along the DWs.

Furthermore, our MFM studies also reveal that there are two DW states [($DW_I$, $DW_{II}$) = (↑,↓) or (↓,↑)] which can be controlled by field cooling sample through $T_N$ with different magnetic field orientations, as shown in fig. 1b and 1c. We define the DWs with dark (bright) color in MFM image, i.e. uncompensated moment parallel (anti-parallel) with the orientation of cooling magnetic field as $DW_I$ ($DW_{II}$). It is possible that $DW_I$ has a larger uncompensated magnetization that $DW_{II}$ around $T_N$ so that it is always parallel to the orientation of cooling field. Unfortunately our MFM results are not sufficient to pin down this possibility. Note that there are two degenerate (anti-)chiral spin states of $Mn^{3+}$ $B_2$ symmetry, which are related to each other by time reversal symmetry (supplementary fig. S5). Two degenerate chiral $Mn^{3+}$ spin states provide a natural origin of the two magnetic DW states [($DW_I$, $DW_{II}$) = (↑,↓) or (↓,↑)] because DM interaction is proportional to the cross product of adjacent spins[18,19]. Therefore, the correlation of uncompensated moments at DWs shown in fig. 2 indicates a single chirality of $Mn^{3+}$ spin state over a large area (18×18 $\mu m^2$), and possibly even over the entire sample. If both chiral states coexist in the field of view, the correlation of uncompensated moments at DWs would be short range (within single chiral domains), as shown in supplementary fig. S6a. In short, within our simple model, the correlation of DW magnetic state indicates a single chiral $Mn^{3+}$ spin state, and field-controlled DW magnetic states indicate field-controlled chiral $Mn^{3+}$ spin states as shown in supplementary fig. S6. Therefore, our simple scenario suggests a potential method of visualize the spin chiral states, which are difficult to detect with conventional neutron scattering methods[29], by harnessing intriguing DW magnetism.

With uniquely correlating ambient PFM and low temperature MFM images, we identified unprecedented collective DW magnetism, i.e. alternating uncompensated $Er^{3+}$ moments at antiphase-ferroelectric DWs interconnected through self-organized vortex network in mutliferroic *h*-$ErMnO_3$. Our analysis indicates that the collective magnetism at multiferroic vortex DWs may stem from a field-controllable chiral $Mn^{3+}$ spin state over the entire sample. Our work not only points to the possibility of probing domain chiral states through DW magnetism, but also demonstrates a new route to realizing nanoscale magnetoelectric coupling, which may be harnessed for multifunctional applications[2,30,31].

**Methods**

Plate-like $ErMnO_3$ single crystals with flat surfaces perpendicular to the *c* axis were grown using a conventional flux method with $Bi_2O_3$ flux. A mixture of 80 mol % of $Bi_2O_3$ and 20 mol % of $ErMnO_3$ powders was heated up to 1260 $^oC$ in a platinum crucible, and cooled slowly with 2.5 $^oC$/h rate. Two pieces from the same batch were heated again up to 1200 $^oC$, and slowly cooled to 1130 $^oC$ with 15 $^oC$/h. One piece is used in SQUID measurement, the other is used in PFM and MFM experiments. The specimen was glued to a sapphire substrate by Ag epoxy. Ambient PFM images were taken on bare (001) surface with a Multimode atomic force microscope in conjunction with homemade circuits. A 50 nm gold film was deposited onto the sample surface with DC sputtering method before MFM experiment to eliminate the electro-static interaction between ferroelectric domains and the MFM tip (fig. 1**d**). MFM images were taken with a home-made VT-MFM, which is interfaced with a Nanoscope IIIa controller. The MFM signal was taken in a frequency-modulated Linear mode, in which topography and MFM scan lines are interleaved. The lift height is 50 nm for all the MFM images except Fig. 2**d** (180 nm to avoid large topography features). Approximately, darker (brighter) MFM contrast indicates more attractive (repulsive) interaction between MFM tip moment and the sample stray field. During MFM experiment, the sample space was kept in cryogenic vacuum for high sensitivity. The alignment between PFM and MFM images is achieved by correlating topographic features.

**Acknowledgements** We thank V. Kiryukhin and Y. Horibe for helpful discussion. The work at Rutgers is supported by NSF grant DMR-0844807 and DMR-1104484.

**Author Contributions** WW and SWC conceived and designed the experiment. SWC and NL grew and annealed ErMnO$_3$ crystals. YJC characterized the magnetic properties. YG carried out PFM and MFM measurements and analyzed the data. WW and SWC wrote the manuscript. YG wrote the supplementary information. All authors contributed to planning and discussions.

**Author information** Reprints and permission information is available at www.nature.com/reprints. The authors declare no competing financial interest. Readers are welcome to comment on the online version of this article at www.nature.com/nature.

Correspondence and requests for materials should be addressed to W.W. (wdwu@physics.rutgers.edu).

---

[*] Present address: Department of Physics and IPAP, Yonsei University, Seoul 120-749, Korea.

## Supplementary Information

## Collective magnetism at multiferroic vortex domain walls

Yanan Geng, N. Lee, Y.J. Choi*, S-W. Cheong and Weida Wu

*Department of Physics and Astronomy and Rutgers Center for emergent materials, Rutgers University, Piscataway, NJ 08854 USA*

**Discussion 1**

Several novel techniques, e.g. x-ray magnetic linear dichroism (XMLD) [1], scanning micro-diffraction[2], and spin polarized-scanning tunneling microscopy (SP-STM) [3], are able to visualize antiferromagnetic domains or walls in very limited cases. None of these state-of-the-art techniques are applicable because $h$-$RE$MnO$_3$ is an insulating antiferromagnet with non-collinear commensurate spin structure. XMLD can only detect orientation domains of collinear antiferromagnets, scanning micro-diffraction can only distinguish domains with different structural modulation wave-vectors, and STM is only applicable for conductive samples. Second harmonic generation (SHG) optics has been shown to be a powerful tool for imaging antiferromagnetic domains in $h$-$RE$MnO$_3$, but it cannot reveal submicron vortex domain pattern because of its relatively poor spatial resolution[4]. Magnetic force microscopy (MFM) has been widely applied in storage industry because it can detect stray field gradient of local moments with nanoscale resolution and good sensitivity[5]. Although advanced MFM has successfully visualized uncompensated moments at the interface of exchange bias systems[6], there is no report of MFM imaging of uncompensated moments in antiferromagnetic DWs.

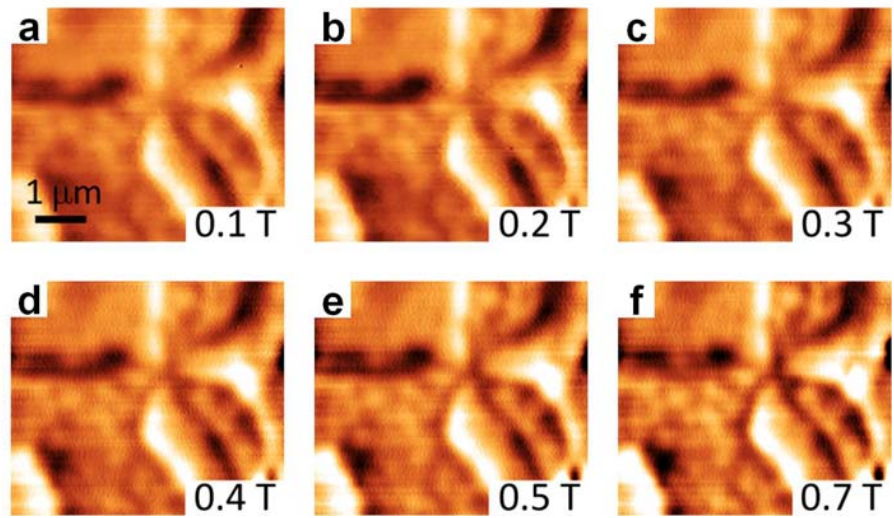

**Figure S1 | magnetic field dependence of MFM images at 5.5 K.** The color scale is 0.8 Hz. Magnetic field values (0.1 – 0.7 T) are labeled at the bottom right corner of each image.

---

* Present address: Department of Physics and IPAP, Yonsei University, Seoul 120-749, Korea

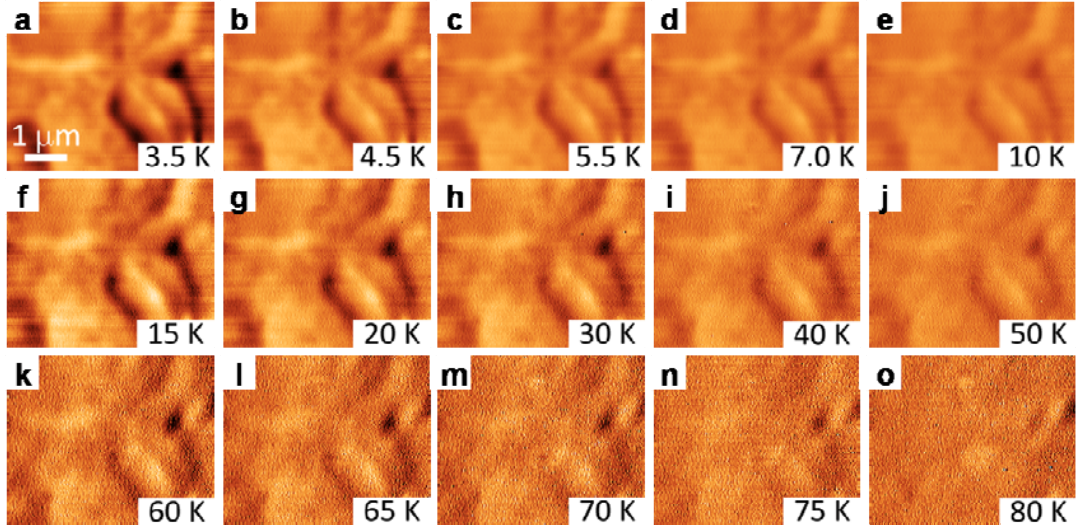

**Figure S2 | Complete set of the temperature dependence of MFM images in 0.2 T.** The color scale ranges for image **a-e**, **f-j**, **k-o** are 2 Hz, 0.5 Hz, 0.2 Hz, respectively.

**Discussion 2**

The characteristic temperature dependence of $RE^{3+}$ moment ($m_{RE}$) is an S-shape curve described by $m_{RE}(T) = m_{RE}(0) \cdot \tanh[\Delta(T)/2k_B T]$, where $\Delta(T) = \Delta_0 \cdot m_{Mn}(T)/m_{Mn}(0)$ is the splitting of doublet due to $Mn^{3+}$ 120° antiferromagnetic order moment $m_{Mn}(T)$. Assuming MFM signal ($\Delta f$) at DWs is proportional to the size of uncompensated magnetization, we obtained a good fit (red curve in fig. 3**g**) of the temperature dependence of DW contrast ($\Delta_0 = 8.7 \pm 1.5$ K, and $T_N = 80 \pm 1$ K) using the empirical results of $m_{Mn}(T) = m_{Mn}(0) \cdot [1-(T/T_N)^d]^e$ from neutron scattering (with fixed parameters: $d \approx 2.36$ and $e \approx 0.31$)[7]. The fitting value of $\Delta_0$ is consistent with the measured values in $HoMnO_3$ (11 K), $YbMnO_3$ (16 K) and $TmMnO_3$ (20 K)[8-10]. Therefore, we conclude that our MFM signal of coupled antiferromagnetic, structural antiphase and ferroelectric DW originates from uncompensated $Er^{3+}$ moments polarized by 120° antiferromagnetic order of $Mn^{3+}$ spins through DM interaction (see discussion 3 for detail).

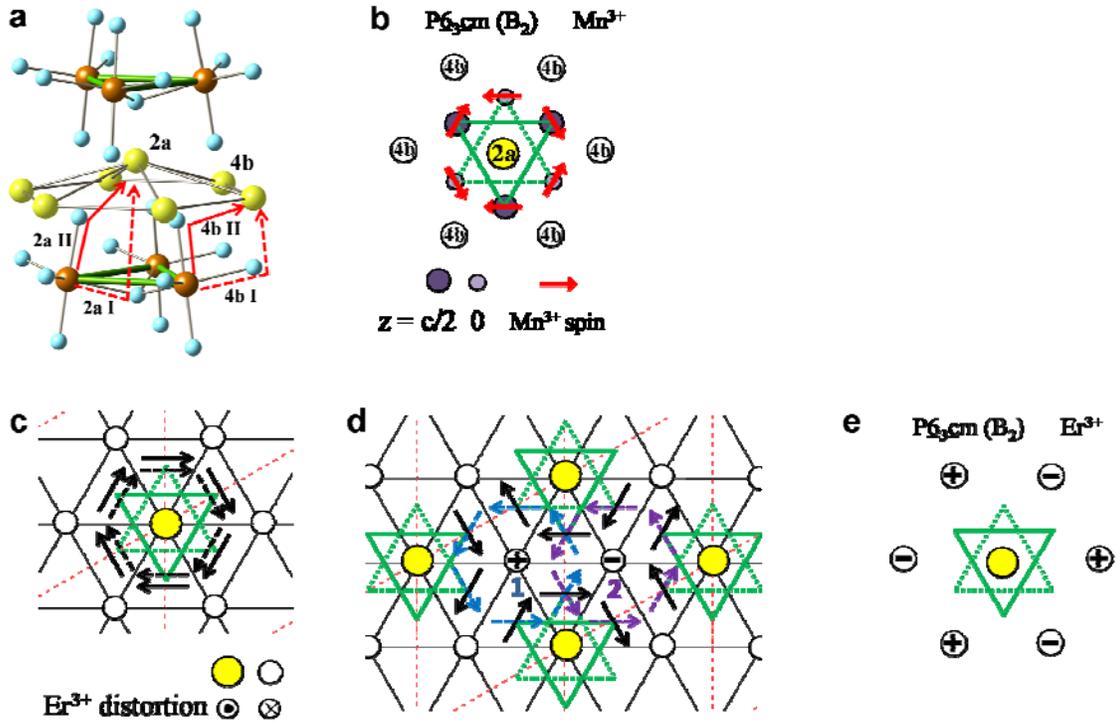

**Figure S3. DM interactions between $Er^{3+}$ and $Mn^{3+}$ in $B_2$ phase within ferroelectric down domain. a**, 3D cartoon of hexagonal $ErMnO_3$ structure ($P6_3cm$), where yellow, brown, light blue spheres represent $Er^{3+}$, $Mn^{3+}$, $O^{2-}$ ions, respectively. The red lines show the DM exchange paths between $Mn^{3+}$ and $Er^{3+}$ at 2a and 4b sites. Path I (II) is defined as interacting through the planar (apical) oxygen, and is noted by dotted (solid) red line. **b**, $Mn^{3+}$ spin configuration of $B_2$ magnetic symmetry. Solid (dotted) green triangles represent $Mn^{3+}$ trimers at z = c/2 (z = 0) layer that is above (below) $Er^{3+}$ layer. $Mn^{3+}$ ions and spins are not shown in the rest cartoons for clarity. Dotted (solid) DM vectors are for path I (II). **c(d)**, DM vectors and induced magnetic moment for $Er^{3+}$ at 2a (4b) site with ferroelectric polarization pointing into paper. In **d**, blue and purple dotted arrows represent DM vectors of the exchange path I for site 4b(1) and 4b(2), respectively. **e**, the induced $Er^{3+}$ AFM order also respects $B_2$ ($P\underline{6}_3\underline{c}m$) symmetry.

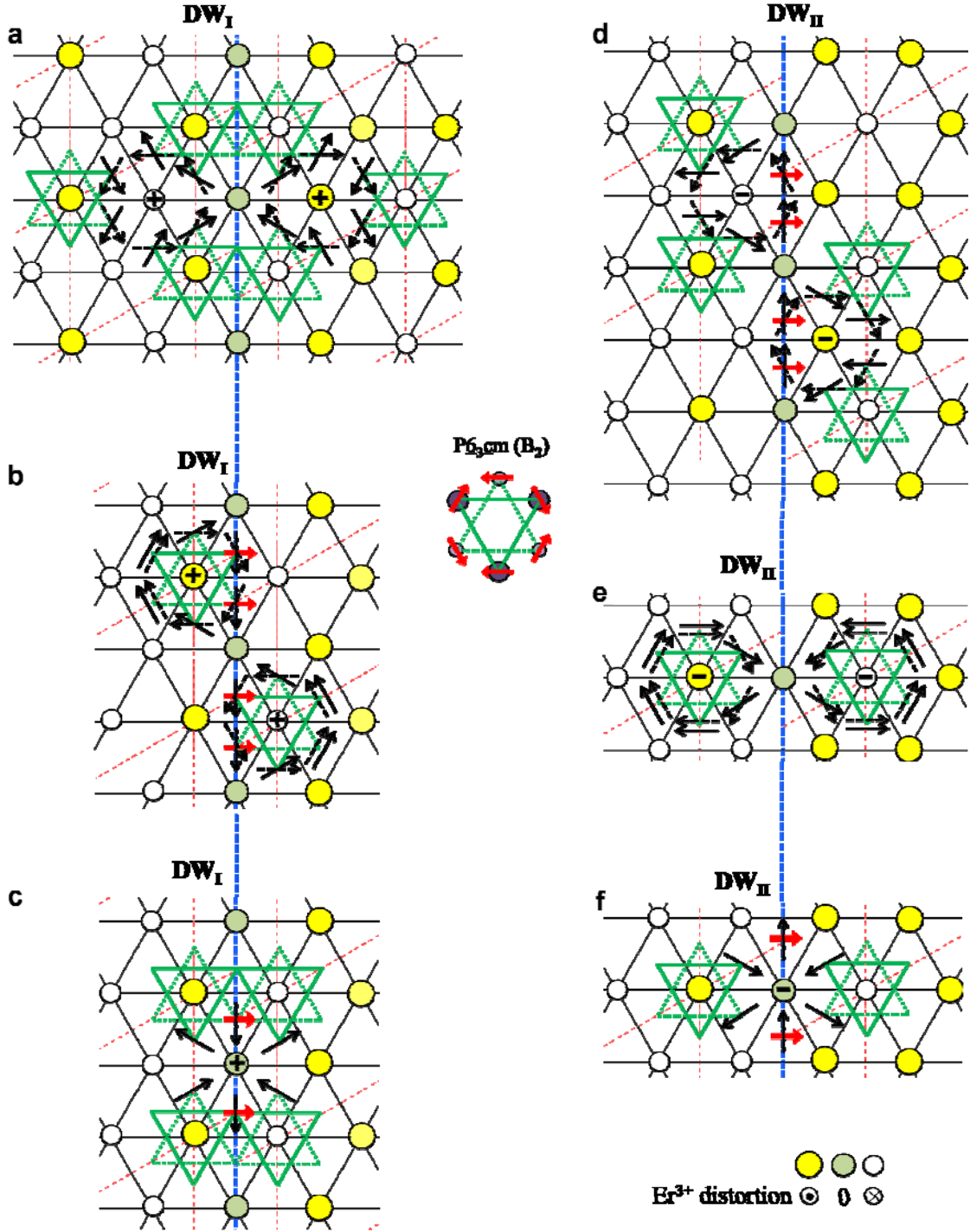

**Figure S4. DM interactions between $Er^{3+}$ and $Mn^{3+}$ in $B_2$ phase along the DWs. a-c (d-f)**, DM vectors and induced magnetic moment for $Er^{3+}$ near or at $DW_I$ ($DW_{II}$). Dotted (solid) DM vectors are for path I (II). We assume the $Mn^{3+}$ spins at the DW plane rotate half way between neighboring domains, as depicted by red arrows (in **b, c, d, f**). For **c** and **f**, no DM vector of path I is present because we assume that there is no lattice distortion at the atomic plane of DW due to antagonistic lattice distortions in neighboring domains.

**Discussion 3: DM vectors**

The DM interactions between an $Er^{3+}$ and nearest neighbor $Mn^{3+}$ moments is

$$H_{DM} = \sum_i \vec{D}_i \bullet (\vec{S}_i \times \vec{\sigma}),$$

where $\vec{D}_i, \vec{S}_i, \vec{\sigma}$ are DM vector, $Mn^{3+}$ spin, $Er^{3+}$ spin, respectively. Using permutation, we can rewrite the DM interactions as:

$$H_{DM} = -\sum_i \vec{\sigma} \bullet (\vec{D}_i \times \vec{S}_i) = -\vec{\sigma} \bullet \vec{H}_{eff},$$

where $\vec{H}_{eff} \equiv \sum_i \vec{D}_i \times \vec{S}_i$ is the effective field that polarizes the $Er^{3+}$ spins. The direction of DM vector is defined as the rotation axis of the distortion of Mn-O bond from high symmetry position with right hand rule[11]. The DM vector magnitudes are denoted as $D_{1a}$ ($D_{1b}$), $D_{2a}$ ($D_{2b}$) for the exchange path I, II of 2a (4b) sites. Using these notations, the effective fields for $Er^{3+}$ moments in domain and DWs can be expressed as follows:

| Domain | $\vec{H}_{eff}$ | $Er^{3+}$ moment | Panel |
|---|---|---|---|
| 2a | 0 | 0 | Fig. **S3c** |
| 4b(1) | $3\sqrt{3}D_{1b}\hat{z}$ | + | Fig. **S3d** |
| 4b(2) | $-3\sqrt{3}D_{1b}\hat{z}$ | - | Fig. **S3d** |

| $DW_I$ | $\vec{H}_{eff}$ | $Er^{3+}$ moment | Panel |
|---|---|---|---|
| 4b(1) | $\left(3\sqrt{2}D_{1b} + D_{2b}\right)\hat{z}$ | + | Fig. **S4a** |
| 2a | $\left(\sqrt{3}D_{1a} + 3D_{2a}\right)\hat{z}$ | + | Fig. **S4b** |
| 4b(DW) | $4D_{2b}\hat{z}$ | + | Fig. **S4c** |

| $DW_{II}$ | $\vec{H}_{eff}$ | $Er^{3+}$ moment | Panel |
|---|---|---|---|
| 4b(2) | $-\left(3\sqrt{3}D_{1b} + 3D_{2b}\right)\hat{z}$ | - | Fig. **S4d** |
| 2a | $-\left(D_{2a}\right)\hat{z}$ | - | Fig. **S4e** |
| 4b(DW) | $-\left(4D_{2b}\right)\hat{z}$ | - | Fig. **S4f** |

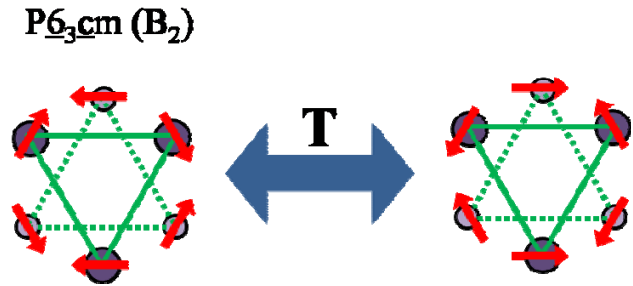

**Figure S5 | Cartoon of two degenerate Mn$^{3+}$ spin states.** The 2 degenerate (anti-)chiral spin states of B$_2$ (*P6$_3$cm*) phase in *h-RE*MnO$_3$ are related to each other by time reversal symmetry.

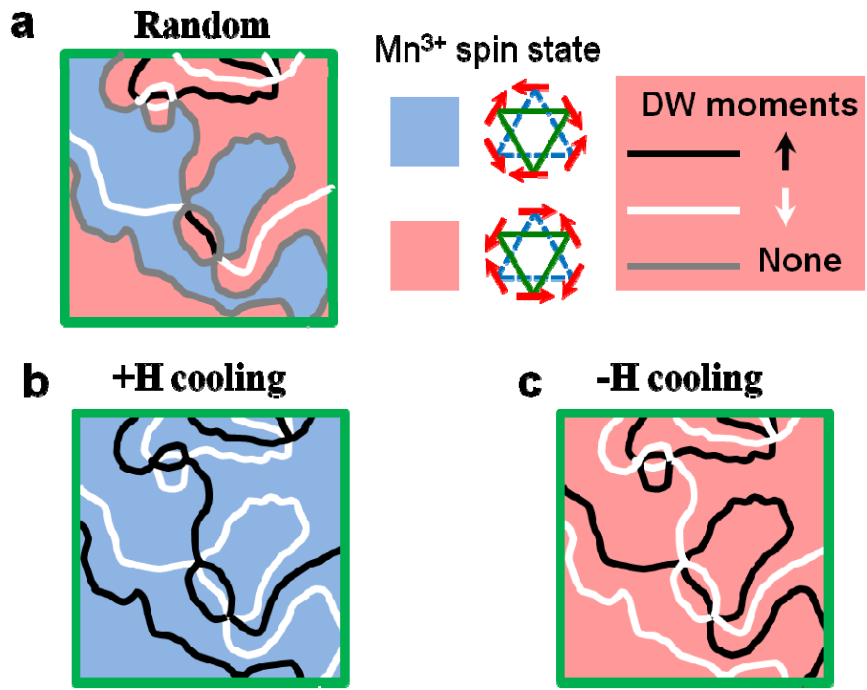

**Figure S6 | Cartoon summary of DW states and chirality of Mn$^{3+}$ spins in *h-RE*MnO$_3$.** In the cartoons, red and blue represent 2 chiral states of Mn$^{3+}$ spins in B$_2$ phase. Black, white and grey lines correspond to up, down and null uncompensated moments at DWs. **a**, DW magnetism of hypothetically random mixing of 2 chiral states. This state is not observed in our MFM experiment. **b** (**c**), single chiral state with alternating DW uncompensated magnetization is obtained after positive (negative) field cooling.